\documentclass[ showpacs,aps]{revtex4-1} 
\usepackage{amsmath}
\usepackage{graphicx} 
\usepackage{epstopdf}       

\begin{document}

\title{\bf Models of spherical shells as sources of   Majumdar-Papapetrou type  spacetimes }


\author{Gonzalo Garc\'{\i}a-Reyes}
\email[e-mail: ]{ggarcia@utp.edu.co}
\affiliation{Departamento de F\'{\i}sica, Universidad Tecnol\'ogica de Pereira,
 A. A. 97, Pereira, Colombia}

\begin{abstract}
By starting with a seed  Newtonian potential-density pair  we construct 
relativistic thick spherical  shell models   for a Majumdar-Papapetrou type  conformastatic  spacetime. As a simple example, we considerer
a family of  Plummer-Hernquist  type  relativistic spherical shells.
As a second application, these  structures   are then used to model a system 
composite by a dust disk and a halo of  matter. 
 We study  the equatorial circular motion of test particles around such  configurations.  Also the stability of the orbits is analyzed for radial perturbation using an extension of the Rayleigh criterion. The models considered   satisfying all the energy conditions. 

\end{abstract}

\maketitle

\section{Introduction}
 Spherically symmetric  distribution of matter  are important 
 in astrophysics as models of  globular clusters, elliptical  galaxies, galactic   bulges  and 
dark matter haloes.  
Spherical shells models  are also useful tools  in astrophysics
and general relativity  as sources of vacuum gravitational fields,  cosmological models, gravitational collapse,  and supernovae  \cite{malafa1, malafa2}.
The pioneering work on relativistic shells is due to Israel \cite{IS1}, who also
specialized his general framework to spherical symmetry  \cite{IS2}. 
 Relativistic spherical shells with varying thickness were    studied 
in \cite{shell-lete1, shell-lete2, Nguyen} for a Schwarzschild type conformastatic spacetime. 

Thin shells  representing the
field  of a  disk are  of astrophysical
importance in  modeling  galaxies, accretion
disks, and the  superposition of a black hole and a galaxy.
 Static thin disks without radial pressure were first studied by  Bonnor and Sackfield \cite{BS},
and Morgan and Morgan \cite{MM1}, and with radial  pressure also by Morgan and Morgan
\cite{MM2}.  
Several classes of exact solutions of the  Einstein field  equations
corresponding to static thin disks with or  without radial pressure have been
obtained by different authors
\cite{LP,CHGS,LO,LEM,BLK,BLP,GE}.
Rotating  thin disks that can be considered as a source of a Kerr metric were
presented by  Bi\u{c}\'ak and  Ledvinka \cite{BL}, while rotating disks with
heat flow were were studied by Gonz\'alez and Letelier \cite{GL2}.
The exact superposition of a disk and a static black hole was first considered
by Lemos and Letelier in Refs. \cite{LL1,LL2}.  Disk-like matter distribution  surrounded by
a material halo for corformastatic spacetimes  have been studied in 
\cite{let-halos} and   for  Einstein-Maxwell fields  in 
\cite{AN-GUI-QUE}  and  \cite{Gutie}. Axisymmetric  conformastatic
disk-haloes were also  considered   in \cite{pimentel} from  solutions of Laplace's 
equation. 

In this work,  we construct models of spherically symmetric  thick shells for a Majumdar-Papapetrou type conformastatic   spacetime
from  a    Newtonian potential-density pair (Poisson's  equation).  
The method  used    allows   to model    different  spherically symmetric  astrophysical systems   such as   such as galactic nuclei and certain star clusters
where relativistic effects are expected to have a significant
impact.  As a simple example,  we consider
a family of  Plummer-Hernquist  type  relativistic spherical shells.  Besides shells, the potential-density pair considered    permits to model  configurations of matter
with a central cusp  and with a finite central density. 
As a second application, these structures   are then used to model
dust disk  surrounded of haloes of  matter. 
The  distributions of matter  studied    are essentially of infinite extension.
However, since the  energy density  decreases rapidly one can define
a cut off radius and,  in principle, to consider
these  objects  as finite.  

The paper is organized as follows. In Sect. II we present the method  to construct  different  spherical configurations of matter from  a Newtonian potential-density pair in the particular case of  
 a  Majumdar-Papapetrou type   conformastatic spacetime. 
In Sect. III the formalism is employed  to build    a  family  of Plummer type relativistic spherical shell,  and  then in  Sect. IV  we model  from them a system 
composite by a dust disk and a halo  of  matter.    In each case 
we  study  the equatorial circular motion of test particles around the
structures.  Also the stability of the orbits is analyzed for radial perturbation using an extension of the Rayleigh criterion \cite{RAYL,FLU}.  Finally, in Sect. V we summarize and discuss the results
obtained.


\section{Majumdar-Papapetrou type  fields and spherical structures    } 
We consider a conformastatic spacetime \cite{synge, kramer} in spherical coordinates 
 ($t$, $r$, $\theta$, $\varphi$) and in the particular form
\begin{equation}
ds^2 =  -f dt^2 + f^{ -1} (dr^2 + r^2d\Omega),  \label{eq:met1}
\end{equation}
where $f(r)$ and  $d\Omega = d\theta^2 +  \sin ^2 \theta  d\varphi ^2$.  These  fields can be called  Majumdar-Papapetrou type  fields  due to its known use  in context of electrostatic fields  \cite{majum, papa}.  
Making  $f = (1- \Phi/2)^{-4} $, with $\Phi(r)$, the same metric takes the form
\begin{equation}
ds^2 =  - \left( 1- \frac{\Phi}{2} \right)^{-4} dt^2 + \left( 1- \frac{\Phi}{2} \right)^{4} (dr^2 + r^2d\Omega).  \label{eq:met}
\end{equation}

The Einstein field equations $G_{ab}= 8 \pi G  T_{ab} $ yield
the following non-zero  components of the energy-momentum tensor  
\begin{subequations}\begin{eqnarray}
T^t \ _t  & = & - \frac{\nabla ^2 \Phi }{4 \pi G   \left( 1- \frac{\Phi}{2} \right)^{5} },
   \label{eq: Ttt}  \\
&  &  \nonumber \\
T^\theta \ _\theta  & = &  T^\varphi \ _\varphi=   \frac{\nabla \Phi \cdot  \nabla \Phi}{8 \pi G \left( 1- \frac{\Phi}{2} \right)^{6}}, \\
& &  \nonumber  \\
T^r \ _r  & = &   - \frac{\nabla \Phi \cdot  \nabla \Phi}{8 \pi G \left( 1- \frac{\Phi}{2} \right)^{6}}.   \label{eq: Trr}
 \end{eqnarray} \end{subequations} 

In terms of the orthonormal  tetrad ${{\rm e}_{ (a)}}^b = \{
V^b , X^b,Y^b,  Z^b \}$, where
\begin{subequations}\begin{eqnarray}
V^a &=& f^{- 1/2}    \delta ^a_t  , \quad  
X^a  = f^{1/2}  \delta ^a_r , \quad   \label{eq:tetrad1} \\
Y^a  &= &   \frac{ f^{1/2}} {r}  \delta ^a_\theta ,  \	 \quad
Z^a =\frac{  f^{1/2} }{ r \sin  \theta }   \delta ^a_\varphi  , \label{eq:tetrad2}
\end{eqnarray}\end{subequations}
the   energy-momentum tensor   can be written as 
\begin{equation}
T^{ab} =  \rho  V^a V^b + p_r X^a X^b + p_\theta ( Y^a Y^b + Z^a Z^b) , \label{eq:perfect}  
\end{equation}
so that the energy density is given by $\rho= - T^t \ _t  $, and the stresses (pressure or tensions)  by  $p_i = T^i \ _i $. 
In order to have a physically meaningful matter distribution the components of the energy-momentum tensor must  satisfy the energy conditions. The weak energy condition
requires that   $\rho\geq 0$, whereas  the dominant energy condition states that 
$|\rho|  \geq |p_i|$. The strong energy
condition  imposes
the condition  
$\rho_{eff} = \rho+ p_r + p_\theta + p_\varphi \geq 0$,  where $\rho_{eff}$
is the ``effective Newtonian density''. 

The metric function $\Phi$   can be chosen by requiring that
in the Newtonian limite  $\Phi \ll 1$  the expression for the
relativistic energy density  must reduce to Poisson's
equation
\begin{equation}
 \nabla ^2 \Phi_N = 4 \pi G \rho_N.  \label{eq: poisson}
\end{equation}
This condition is satisfied by taking  $ \Phi =  \Phi_N$.

Thus,    the physical quantities associated with the matter distribution are given by  
\begin{subequations}\begin{eqnarray}
\rho& = & \frac{ \rho_N } { \left (1- \frac{\Phi_N}{2} \right )^{5} }    ,     \label{eq: en} \\ 
&  & \nonumber   \\
p_\theta & = & p_\varphi =  -  p_r =  \frac{\nabla \Phi_N \cdot  \nabla \Phi_N}{8 \pi G \left( 1- \frac{\Phi_N}{2} \right)^{6}},  \\
&  &  \nonumber
\end{eqnarray} \end{subequations} 
and  the average pressure  by  
\begin{equation} 
p= \frac 1 3 (p_r  + p_\theta + p_\varphi ) =  \frac{\nabla \Phi_N \cdot  \nabla \Phi_N}{24 \pi G \left( 1- \frac{\Phi_N}{2} \right)^{6}} .
\end{equation}

In conclusion, given a seed Newtonian potential-density pair ($\Phi_N$, $\rho_N$) we can constructed for conformastatic fields different models  of spherical  compact  objects   such as spheres,  spherical shells, disks and haloes.  

\subsection{Motion of   particles and stability    } 

We will now analyze the motion of  test particles   around the structures on  the plane $\theta =\pi /2$.
For equatorial circular orbits the 4-velocity  $u^a$  of the particles with respect to the coordinates
frame has  components $u^a = u^t(1,0, 0,\omega )$ where  $\omega= u^\varphi/u^t=\frac{d
\varphi}{d t}$ is  the  angular speed of the test particles.  For the spacetime  (\ref{eq:met}),
the equation for the geodesic motion of the particle is given by
\begin{equation}
 g_{ab, r}u^a u^b = 0,
\label{eq:geo}
\end{equation}
from which we get
\begin{equation}
\omega ^2 = - \frac{g_{t t, r}}{g_{\varphi \varphi, r }}.
\end{equation}
On the other hand,  $u^t$ obtains normalizing  $u^a$, that is requiring  $g_{ab}u^au^b=-1$, so that
\begin{equation}
 (u^t)^2 = - \frac {1}{g_{\varphi \varphi} \omega^2 + g_{tt}} \label{eq:u0}.
\end{equation}
With respect to the  orthonormal tetrad (\ref{eq:tetrad1})-(\ref{eq:tetrad2}) the 3-velocity
has components
\begin{equation}
 v^{ (i)}=  \frac { {{\rm e}^{ (i)}}_a u^a } { {{\rm e}^{(0)}}_b u^b } .
\end{equation}
For equatorial circular  orbits  the
only nonvanishing velocity  component is given by
\begin{equation}
 (v^{ (\varphi)})^2 = v_c^2= - \frac{ g_{\varphi \varphi} }{ g_{tt} } \omega ^2 ,  \label{eq:vc2}
\end{equation}
which represents   the circular speed   (rotation profile)   of the particle as seen by an observer at infinity.
Thus we obtain
\begin{equation}
v_c^2 = \frac {r f_{,r}} {2f-r f_{,r} }.
\end{equation}

Another quantity related with the geodesic motion of the particles  is the specific
angular momentum of a particle rotating at a radius $r$, defined as $h =
g_{\varphi\varphi} u^\varphi$. For the plane $\theta =\pi /2$, we have
\begin{equation}
h^2 = \frac{r^2 f^{-1} v_c ^2}{1- v_c ^2} .
\label{eq:moman}
\end{equation}
This quantity can be used to analyze the stability of the particles against radial
perturbations. The condition of stability,
\begin{equation}
\frac{d(h^2)}{dr} \ > \ 0 ,
\end{equation}
is an extension of the Rayleigh criteria of stability of a fluid in rest in a
gravitational field \cite{FLU}.

 For the above  spherical distribution 
these quantities give
\begin{subequations}\begin{eqnarray}
v_c^2 &=& \frac { v_{Nc}^2}  {1- \frac{\Phi_N}{2} - v_{Nc}^2   } ,  \label{eq:vc}  \\
& & \nonumber  \\
h^2 &=& \frac {r^2(1-\frac{\Phi_N}{2})^4 v_{Nc}^2}  {1- \frac{\Phi_N}{2} -2 v_{Nc}^2   },
 \label{eq:h}
\end{eqnarray} \end{subequations} 
where $v_{Nc}^2 = r \Phi_{N,r}$ is the Newtonian circular speed.


\section{Plummer-Hernquist  type  spherical shell models} 
A simple  Newtonian potential-density pair is \cite{shell-lete1}
\begin{subequations}\begin{eqnarray}
\Phi_N &=& - \frac{GM}{(r^n + a^n)^{1/n}}, \label{eq:plu1}  \\
& & \nonumber   \\
\rho_N &=&  \frac {M (n+1) a^n r^{n-2}}{ 4 \pi (r^n + a^n)^{2 + 1/n}},
\label{eq:plu2}
\end{eqnarray} \end{subequations} 
where $a$ is a non-zero constant with the dimension of length  and $n\geq 1$. 
For $n=1$ the potential-density pair  (\ref{eq:plu1}) and (\ref{eq:plu2}) reduces to Hernquist model which has a density profiles with a central cusp and it  has been used to model elliptical galaxies and bulges \cite{Hernquist}.
For $n=2$ we have the  Plummer's spherical model  \cite{Plummer,  Binney} for globular clusters which has a  mass distribution concentrated at center. This  potential-density pair
has also  been used to model  the central region  of our Galaxy  composed by the 
 bulge/stellar-halo and the   inner core   \cite{Flynn}. 
For $n\geq 3$
we have    a  shell-like matter  distribution. The Newtonian circular speed is given by 
\begin{equation}
v_{Nc} ^2 = \frac{G M r ^n} { (r^n + a^n)^{1 + 1/n}}.  \label{eq:vNc}
\end{equation}

In this case the metric function $f$ is given by 
\begin{equation}
f =  \left[ 1+\frac{GM}{2} \left( r^{n} + a^{n} \right) ^{-1/n} \right] ^{-4} ,  
\end{equation}
and the relativistic expressions for the physical quantities main  associated with the matter distributions are 
\begin{subequations}\begin{eqnarray}
 \rho &= &    \frac{ 8 (1+n) \tilde a^n \tilde r^{n-2}  }
 {\pi  G^3 M^2 \xi ^{2- 4/n} (2 \xi ^{1/n} + 1)^5 },  \\   
& & \nonumber  \\
 p &=&  \frac{ 8 \tilde r^{2n-2} \xi ^{4/n -2}  } { 3 \pi  
 G^3 M^2 (2 \xi ^{1/n} +1 )^6 } ,  \\
& & \nonumber  \\
v_c ^2 &=& \frac {2 \tilde r ^n}{   2 \xi ^{1/n +1} + \xi - 2 \tilde r ^n   } ,  \\
& & \nonumber  \\
h ^2 &=&  \frac{ G^2 M^ 2  \tilde r^{2+n} (2 \xi ^{1/n} + 1 )^2  }{2 \xi ^{2/n}
(  2 \xi ^{1/n +1} +\xi  - 4 \tilde r ^n ) },
\end{eqnarray} \end{subequations} 
 where $\tilde r = r /(MG)$, $\tilde a = a /(MG)$, and $\xi = \tilde r^{n}+ \tilde a ^{n} $.  
 
\begin{figure}
$$
\begin{array}{cc}
\includegraphics[width=0.3\textwidth]{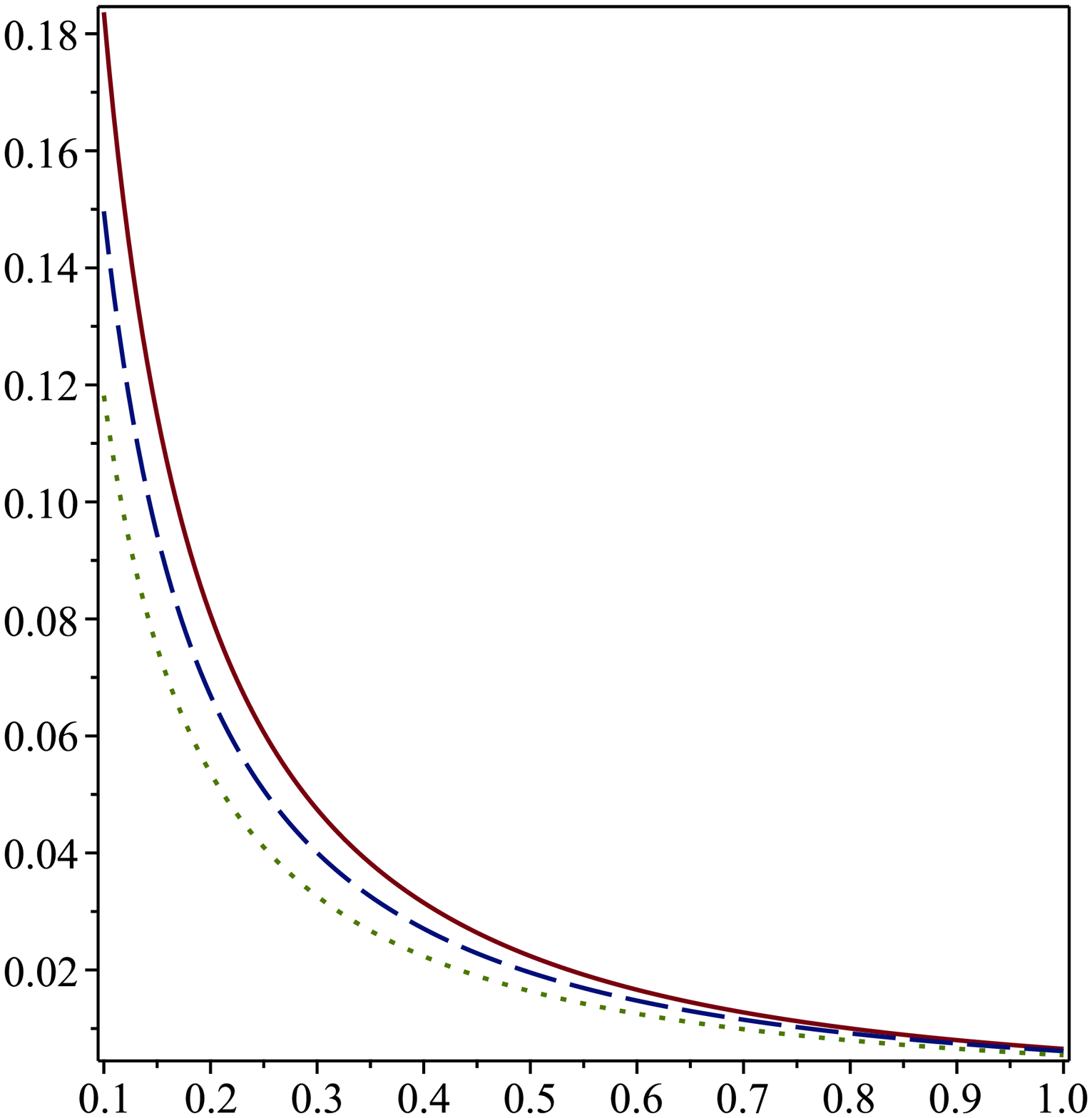} &      
\includegraphics[width=0.3\textwidth]{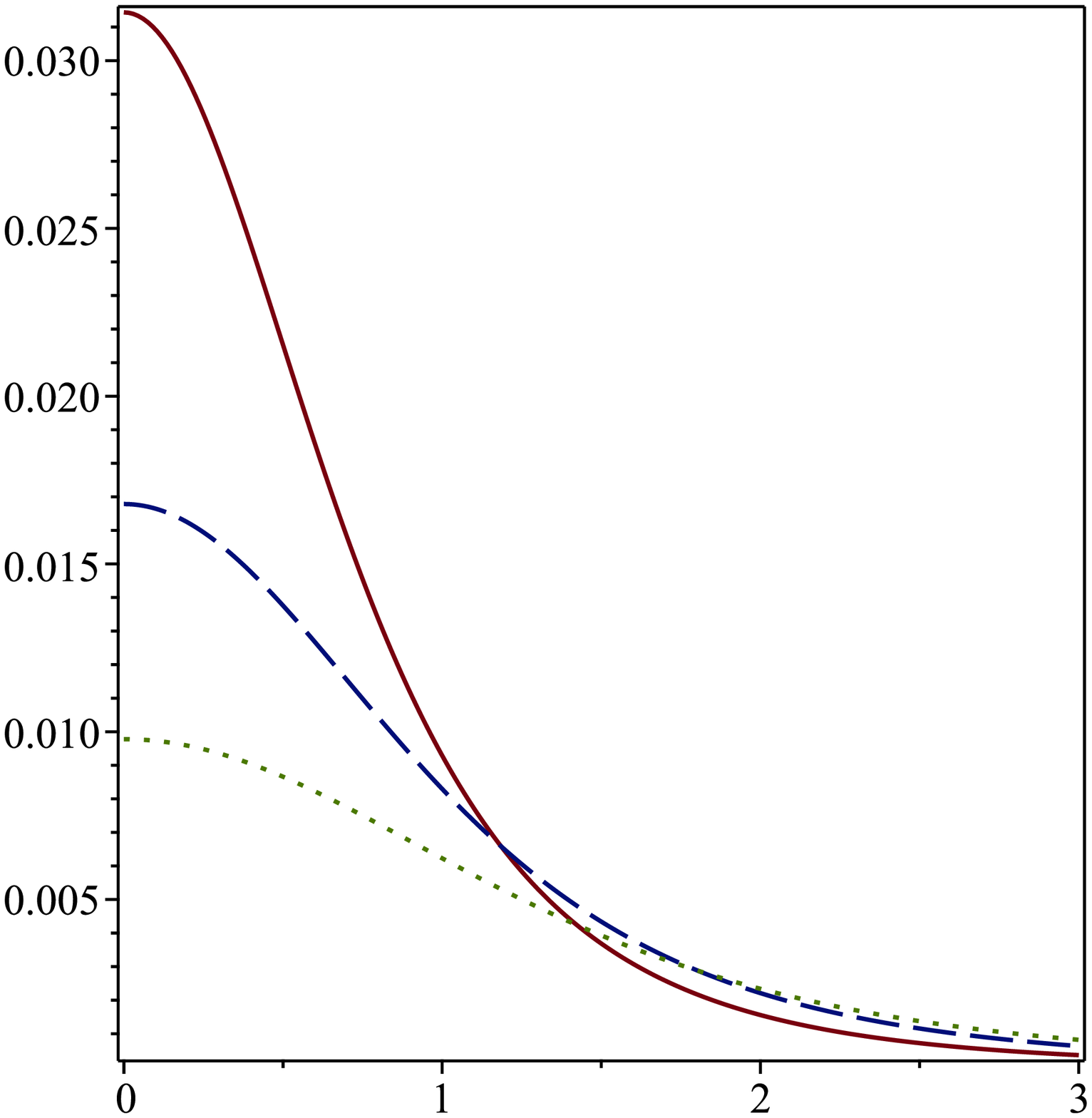}    \\  
 \tilde r & \tilde r  \\
 &  \\
(a)    &  (b)  \\
& \\
\includegraphics[width=0.3\textwidth]{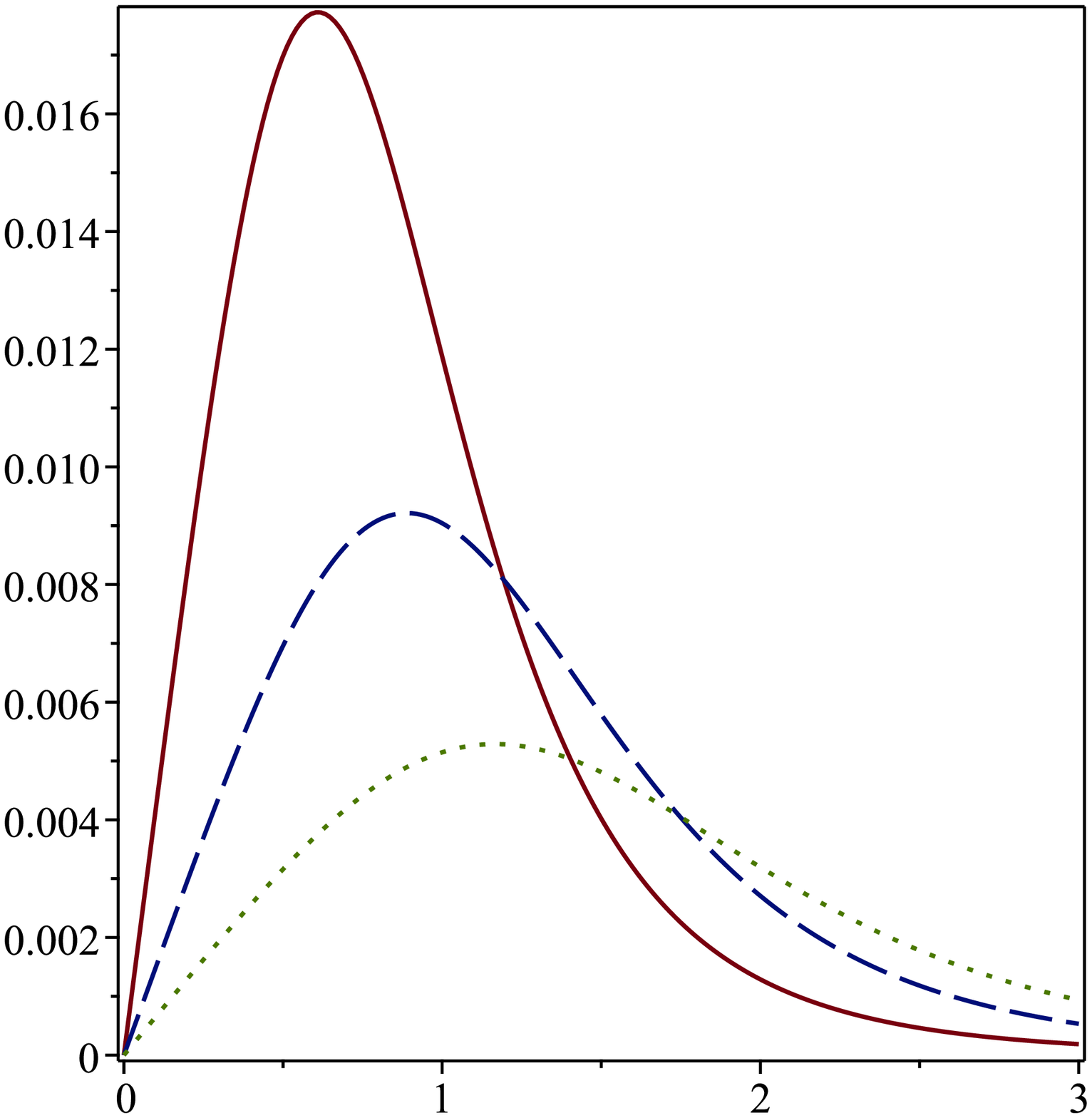} &      
\includegraphics[width=0.3\textwidth]{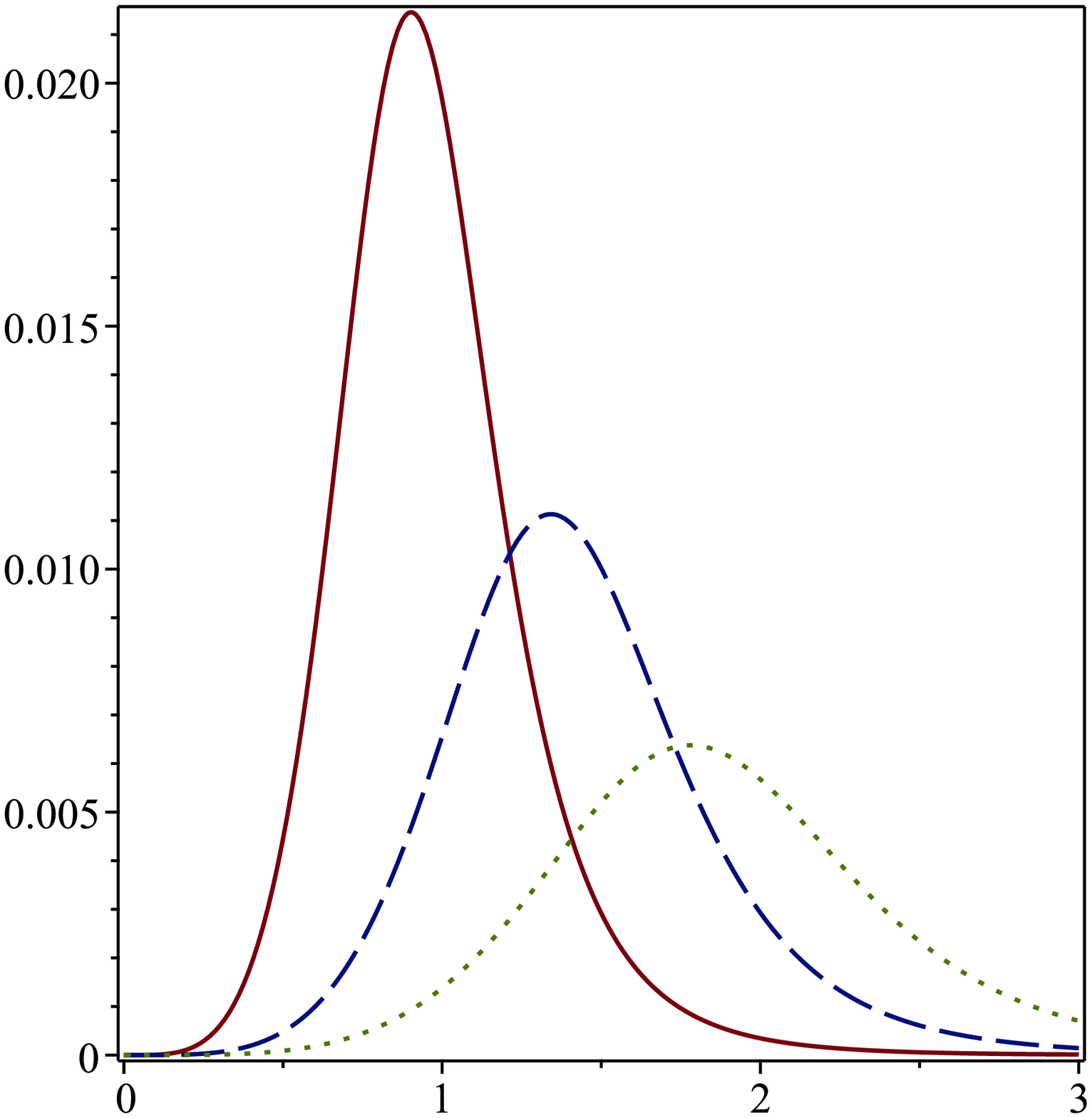} \\
 \tilde r & \tilde r  \\
 &  \\
(c)    &  (d) 
\end{array}
$$	
\caption{ The  relativistic energy density  $\tilde \rho$, as  function of $\tilde r$, 
for  $(a)$ the Hernquist-like model $n=1$, $(b)$ the Plummer-like model $n=2$, 
and the shells $(c)$ $n=3$ and $(d)$ $n=6$, with parameter $\tilde a=1$ (solid curves), $1.5$ 
 (dashed curves)  and $2$ (dotted curves).  }
\label{fig:fig1}
\end{figure}

\begin{figure}
$$
\begin{array}{cc}
\includegraphics[width=0.3\textwidth]{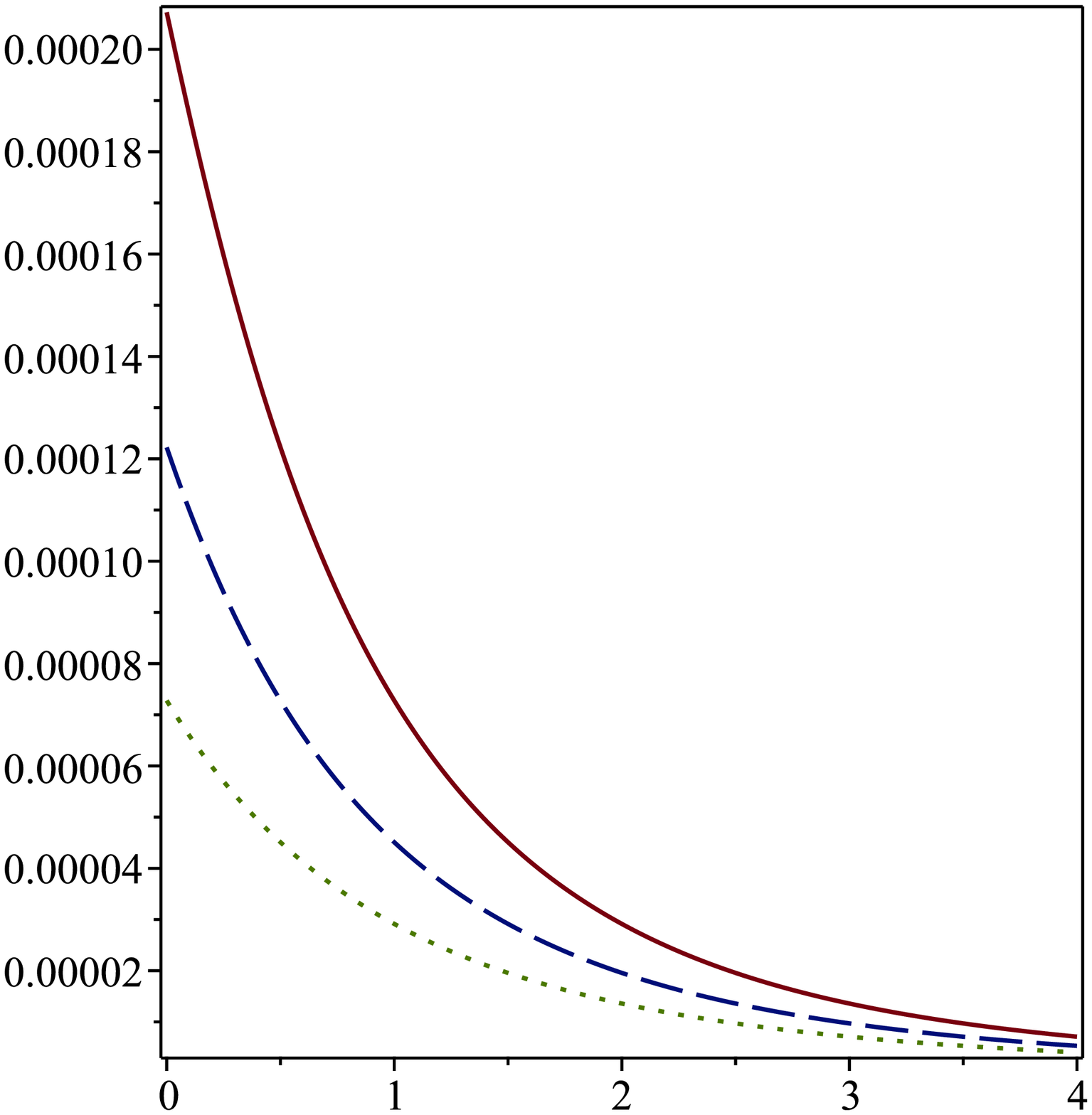} &      
\includegraphics[width=0.3\textwidth]{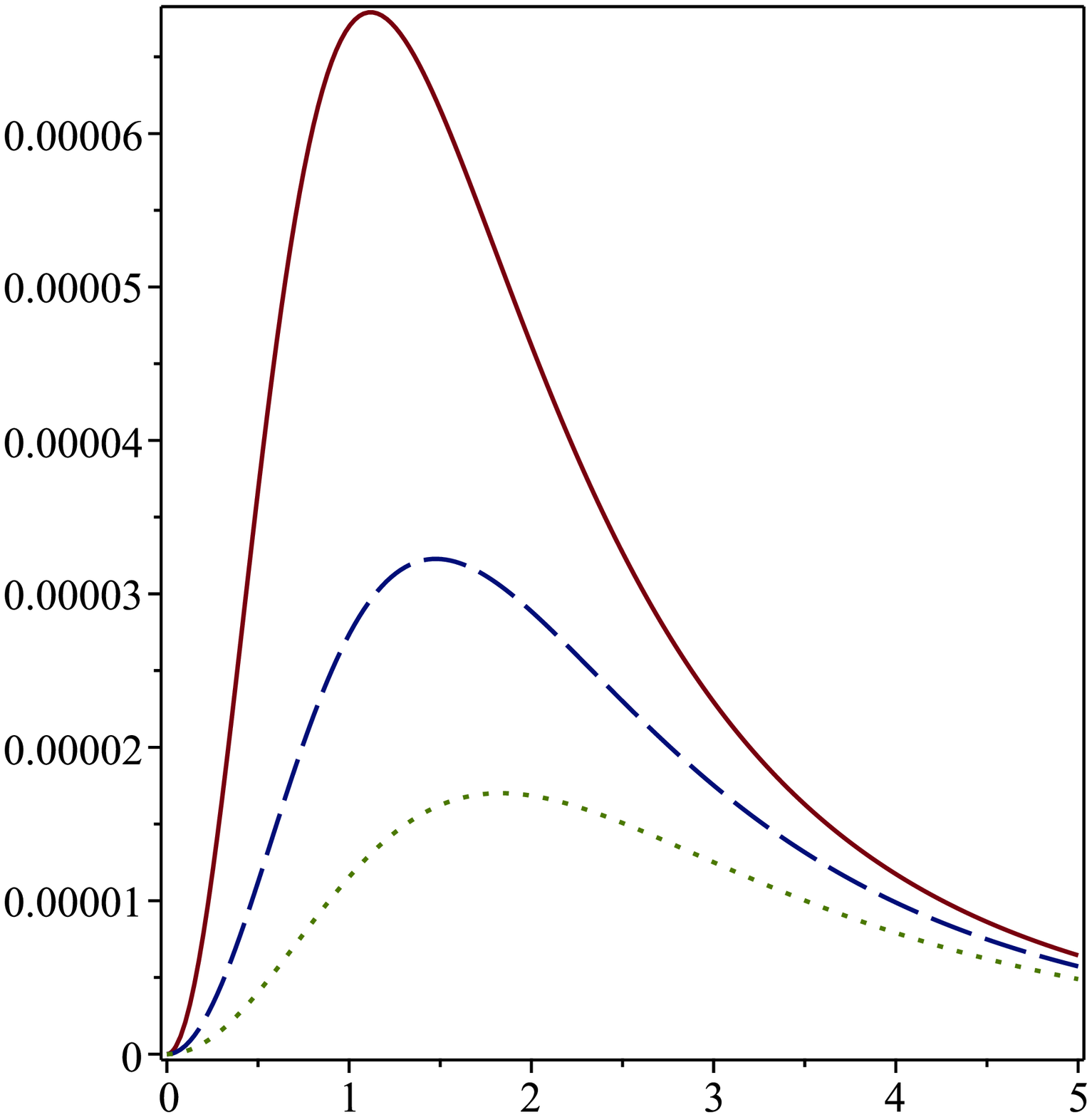}    \\  
 \tilde r & \tilde r  \\
 &  \\
(a)    &  (b)  \\
& \\
\includegraphics[width=0.3\textwidth]{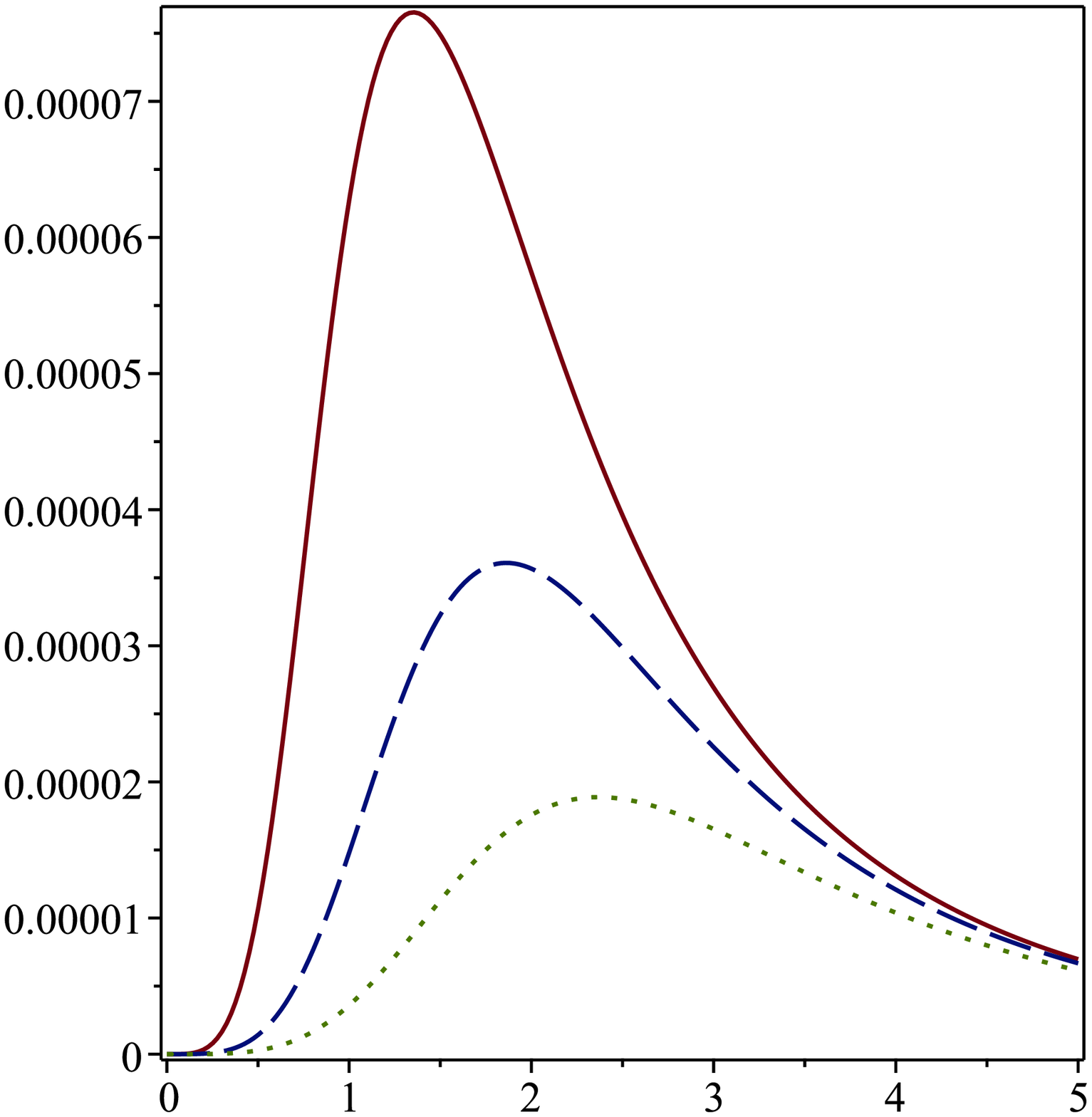} &      
\includegraphics[width=0.3\textwidth]{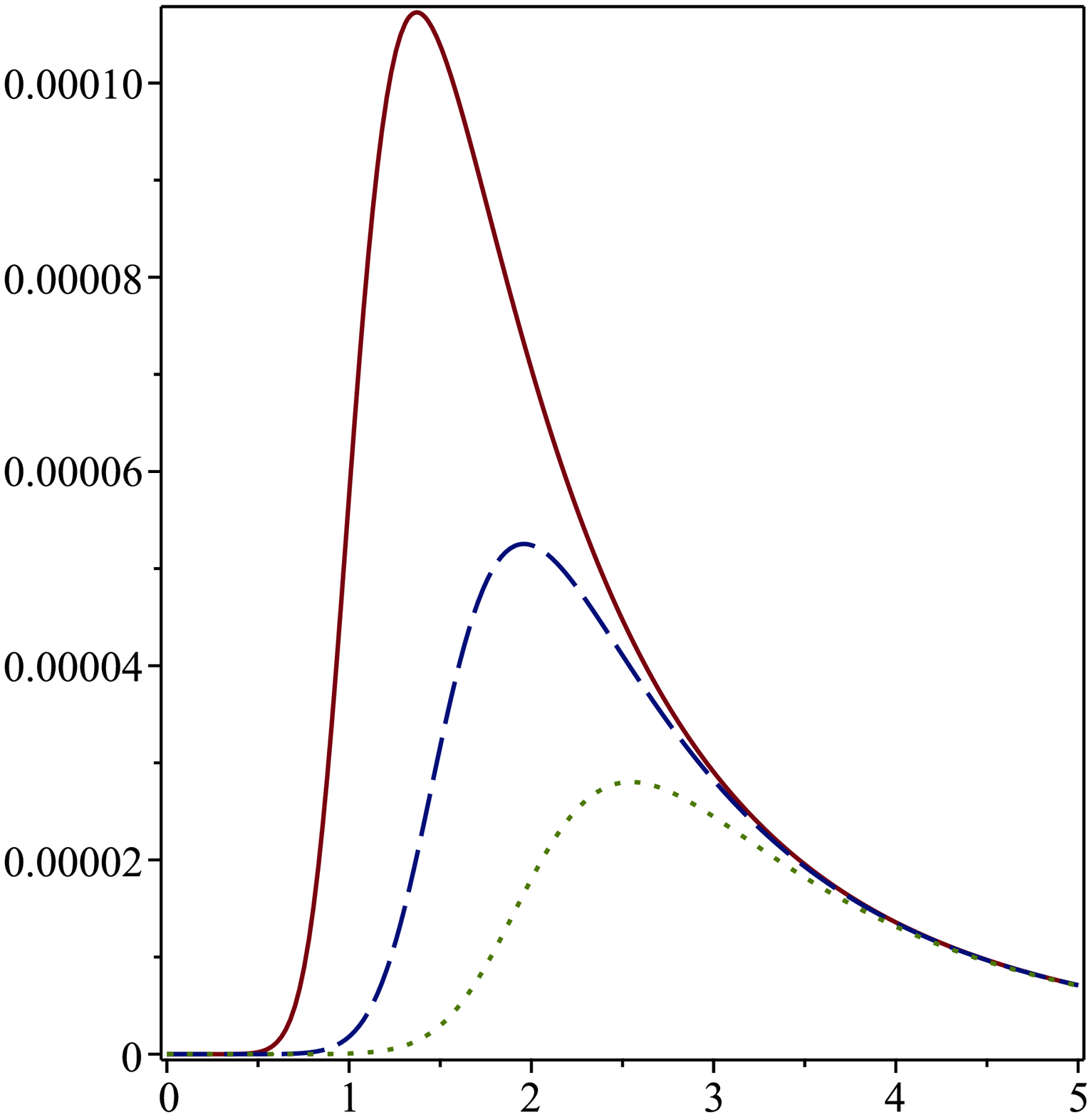} \\
 \tilde r & \tilde r  \\
 &  \\
(c)    &  (d) 
\end{array}
$$	
\caption{The  average pressure $\tilde p$, as  function of $\tilde r$, 
for  $(a)$ the Hernquist-like model $n=1$, $(b)$ the Plummer-like model $n=2$, 
and the shells $(c)$ $n=3$ and $(d)$ $n=6$, with parameter $\tilde a=1$ (solid curves), $1.5$ 
 (dashed curves)  and $2$ (dotted curves).  }
\label{fig:fig2}
\end{figure}

In Figures  \ref{fig:fig1} -  \ref{fig:fig2}  we plot,  as functions of $\tilde r$,  the energy density  $\tilde \rho=  G^3 M^2 \rho$  and the average  pressure $\tilde p=  G^3 M^2  p$   for  the Hernquist-like  model $n=1$,  the Plummer-like  model $n=2$, 
and the  relativistic spherical shells  $n=3$ and  $n=6$, with parameter $\tilde a=1$, $1.5$ 
  and $2$.  When $n=1$ we have a   cusp-like  density profile,
when $n=2$ we have  a  finite central density, and   when $n>2$
the density profile   vanishes  at the origin which suggests  
a shell-like matter  distribution.
In all cases, the energy density  decreases rapidly with $\tilde r$ which  permits
to define a cut off radius $r_c$ and, in principle, to considerer these structures as  compact objects.  The shells   become more concentrated when the parameter $a$ is increased.  We also   find  that the energy is everywhere positive in accordance with the weak energy condition and we have  positive stresses (pressure).
  
\begin{figure}
$$
\begin{array}{cc}
\includegraphics[width=0.3\textwidth]{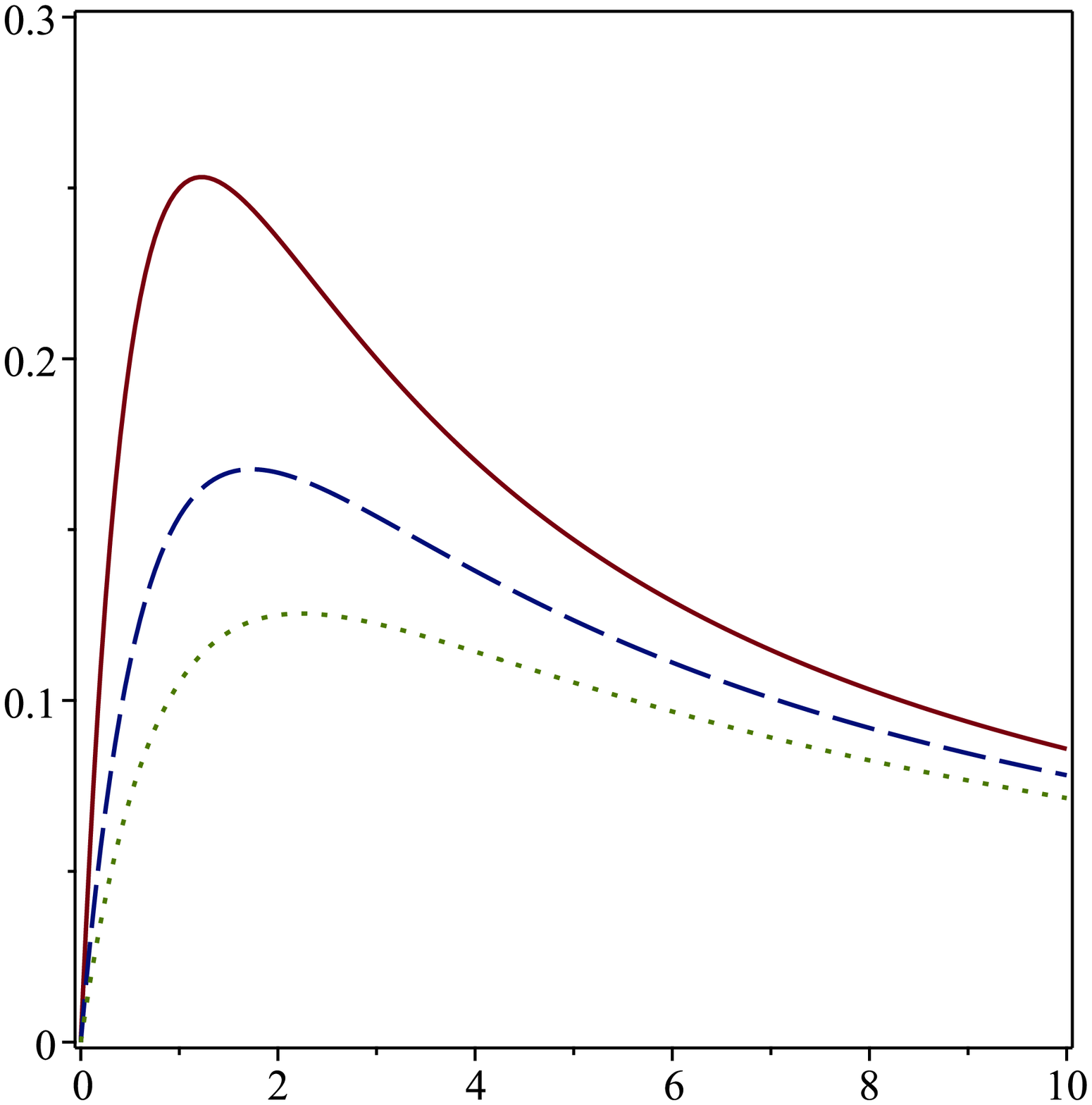} &      
\includegraphics[width=0.3\textwidth]{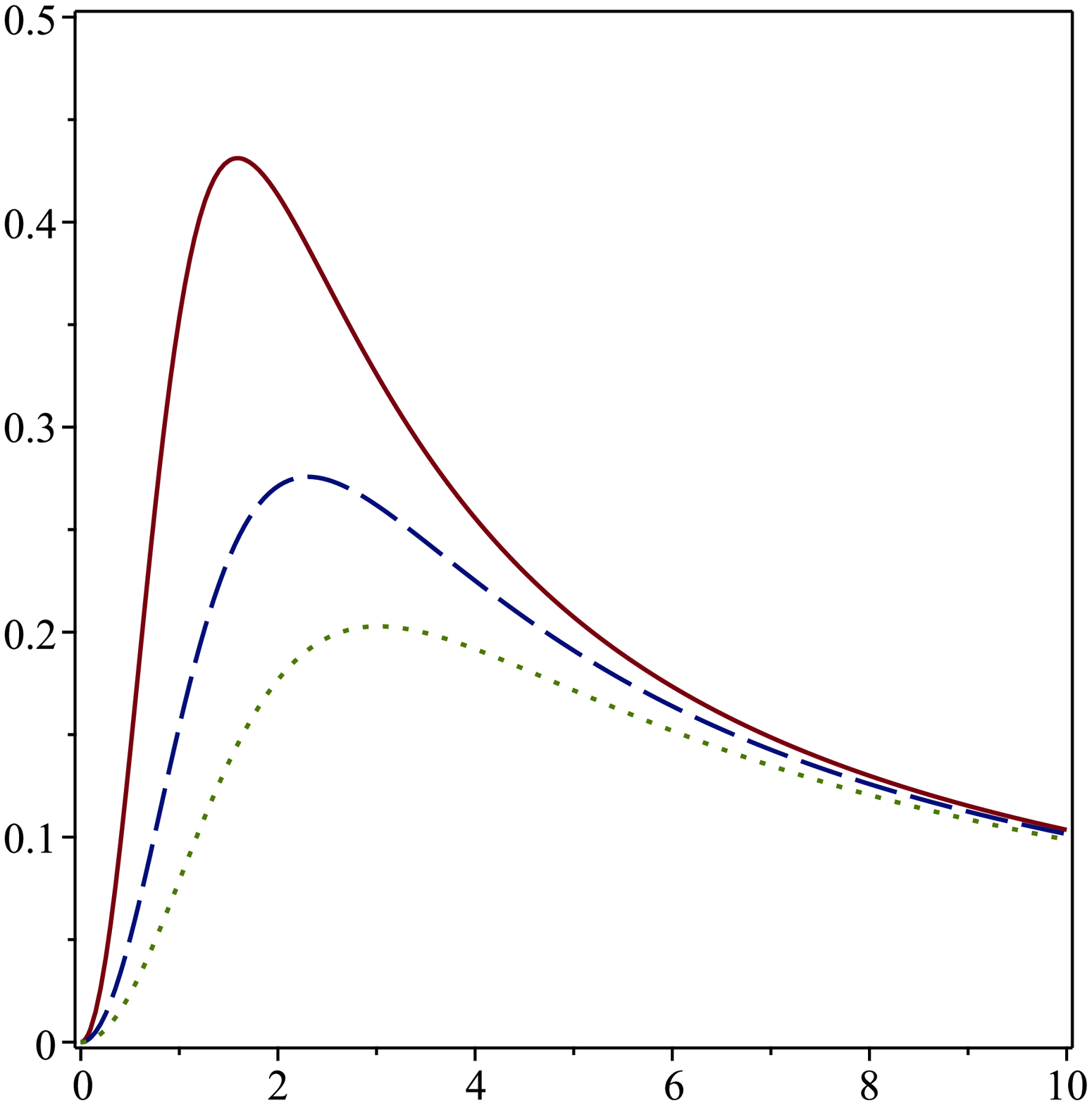}    \\  
 \tilde r & \tilde r  \\
 &  \\
(a)    &  (b)  \\
& \\
\includegraphics[width=0.3\textwidth]{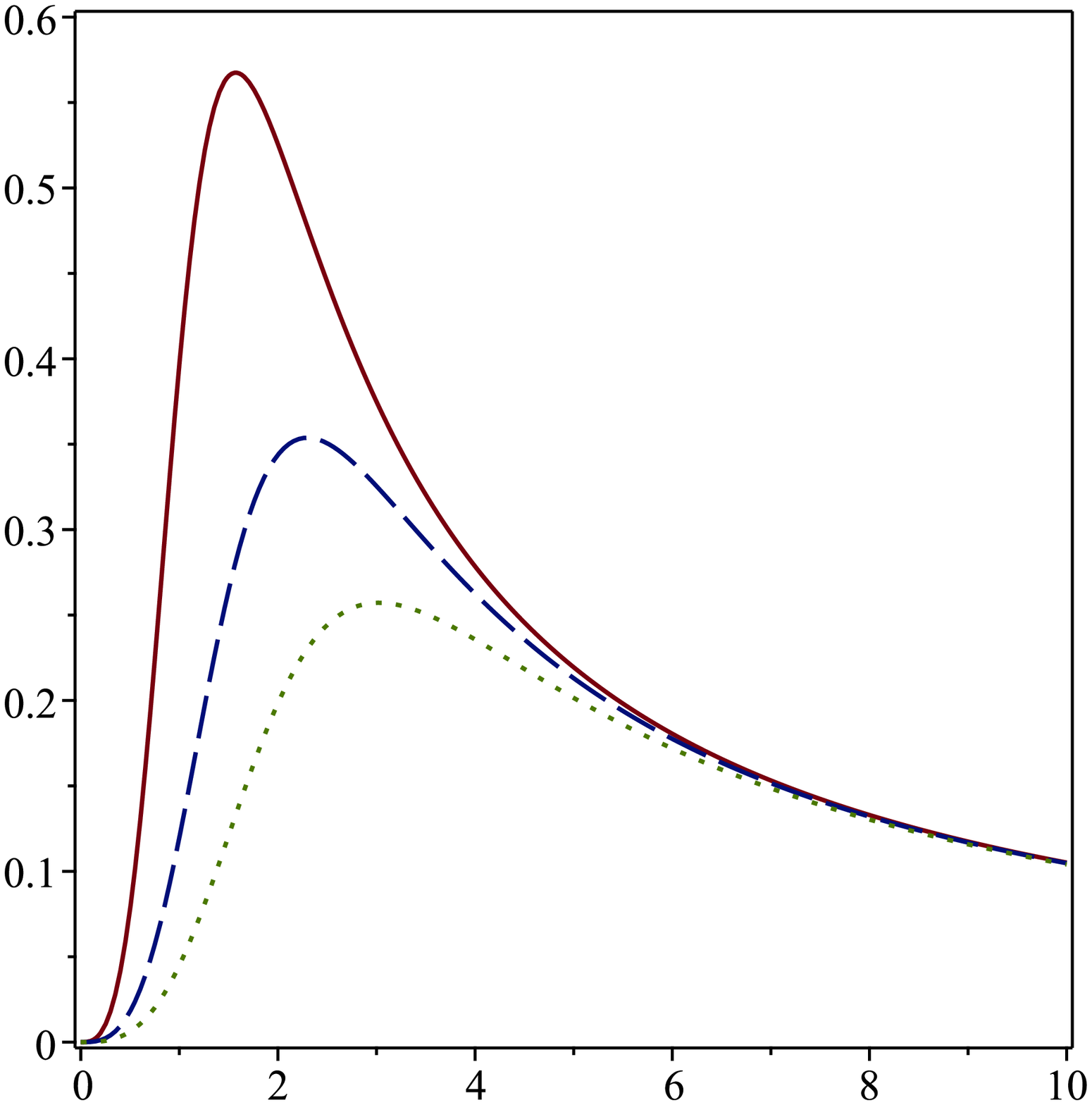} &      
\includegraphics[width=0.3\textwidth]{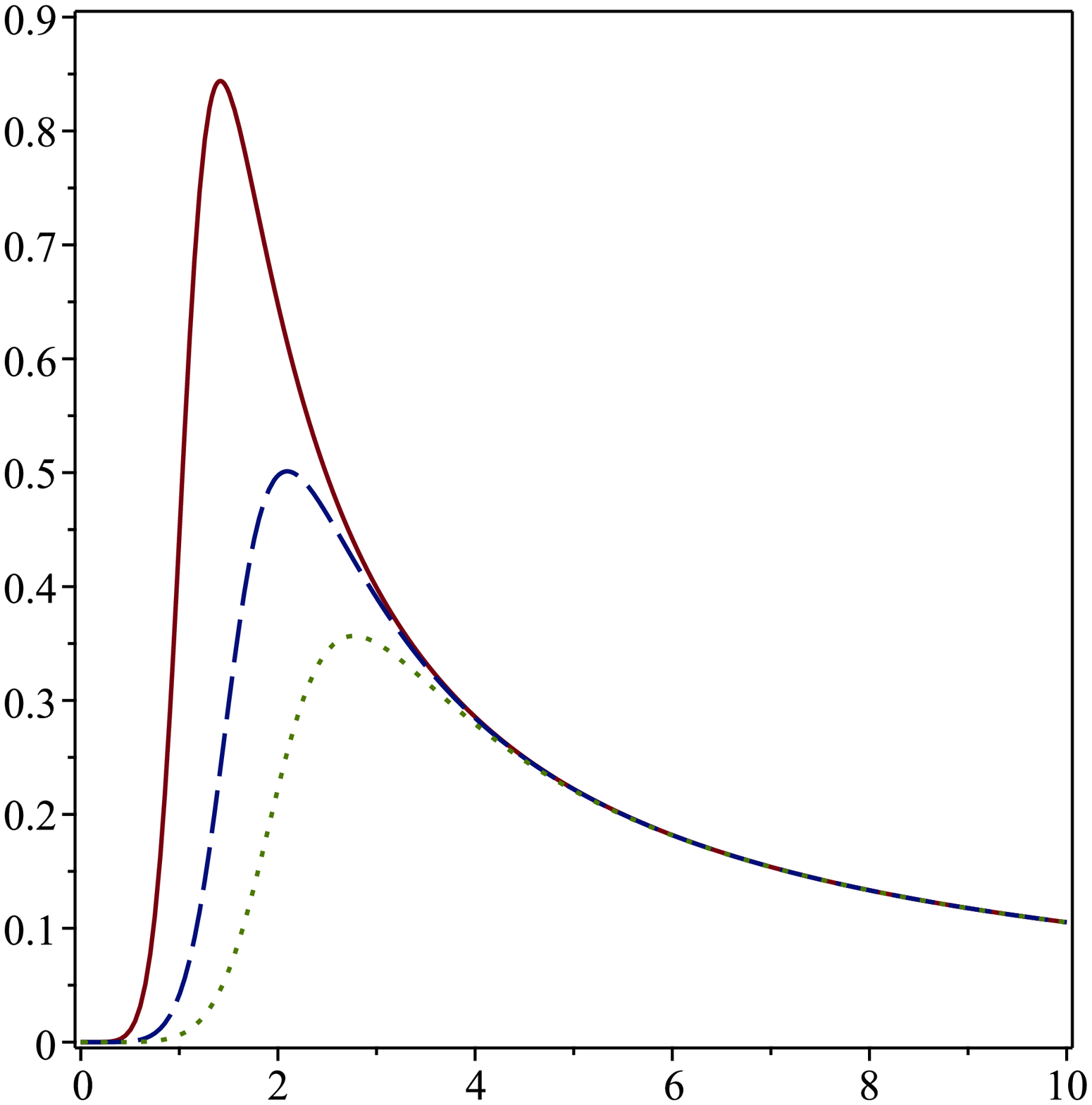} \\
 \tilde r & \tilde r  \\
 &  \\
(c)    &  (d) 
\end{array}
$$	
\caption{ The  rotation curves   $ v_c^2$, as  function of $\tilde r$, 
for  $(a)$ the Hernquist-like model $n=1$, $(b)$ the Plummer-like model $n=2$, 
and the shells $(c)$ $n=3$ and $(d)$ $n=6$, with parameter $\tilde a=1$ (solid curves), $1.5$ 
 (dashed curves)  and $2$ (dotted curves).  }
\label{fig:fig3}
\end{figure}

\begin{figure}
$$
\begin{array}{cc}
\includegraphics[width=0.3\textwidth]{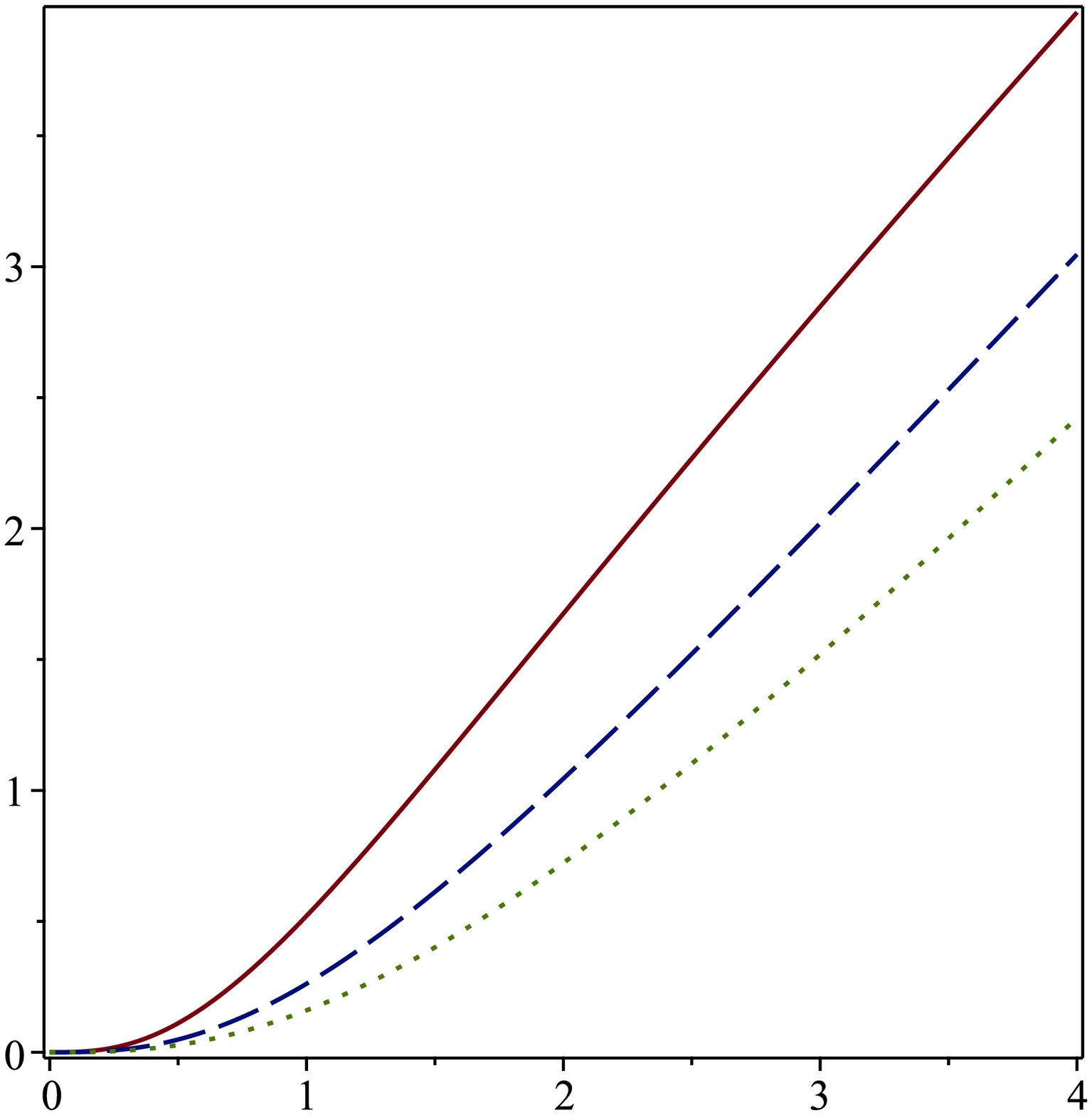} &      
\includegraphics[width=0.3\textwidth]{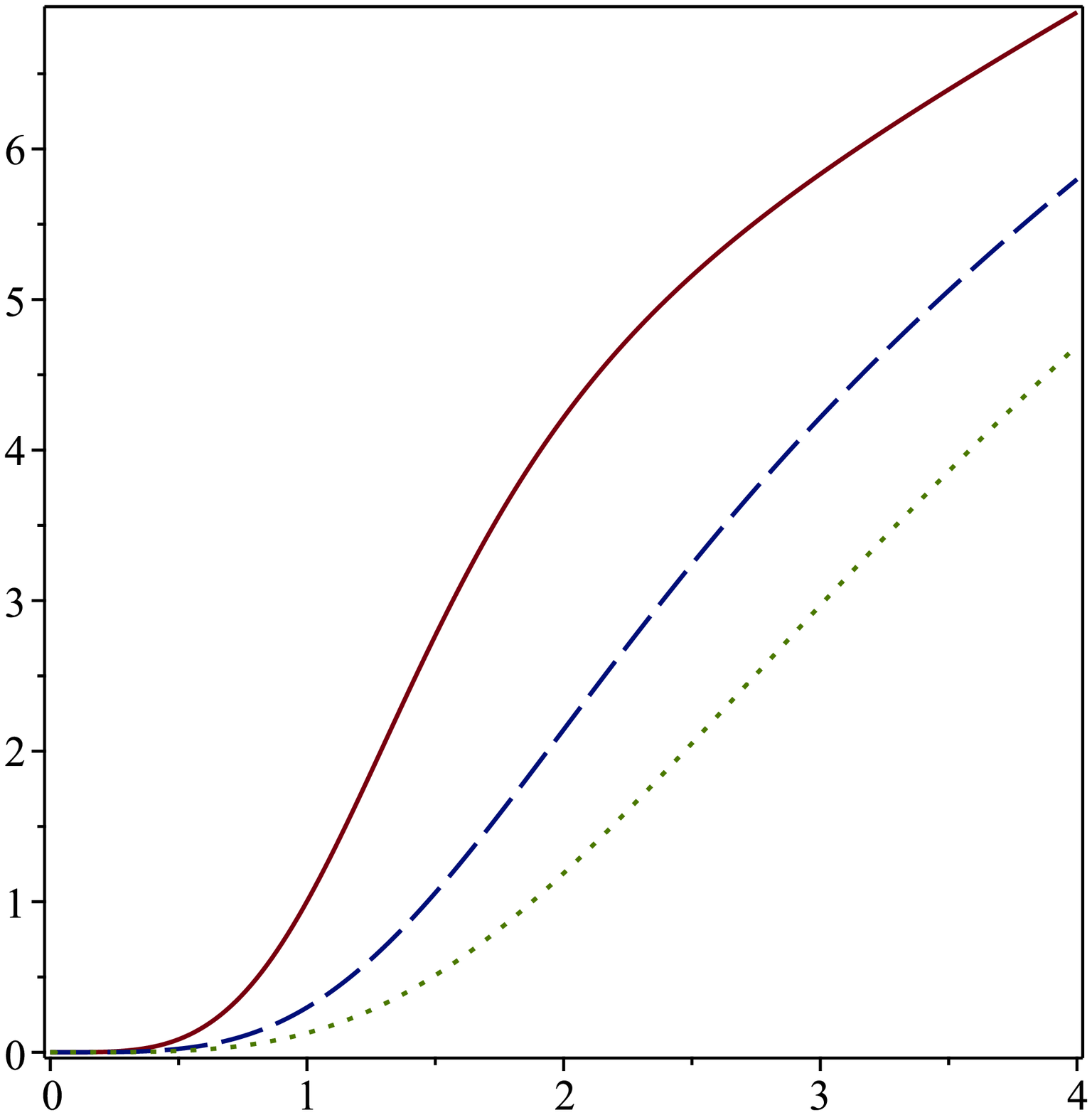}    \\  
 \tilde r & \tilde r  \\
 &  \\
(a)    &  (b)  \\
& \\
\includegraphics[width=0.3\textwidth]{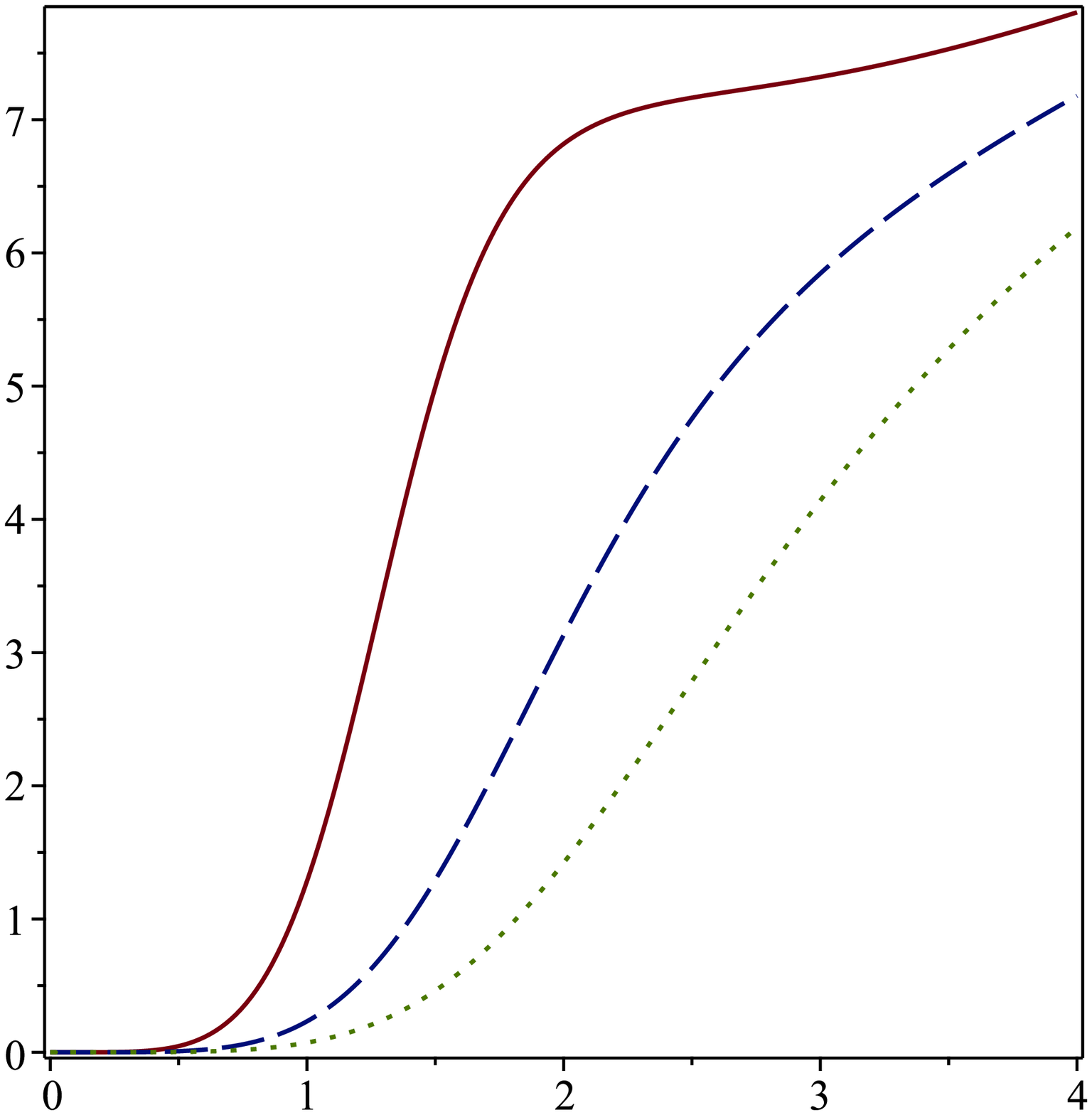} &      
\includegraphics[width=0.3\textwidth]{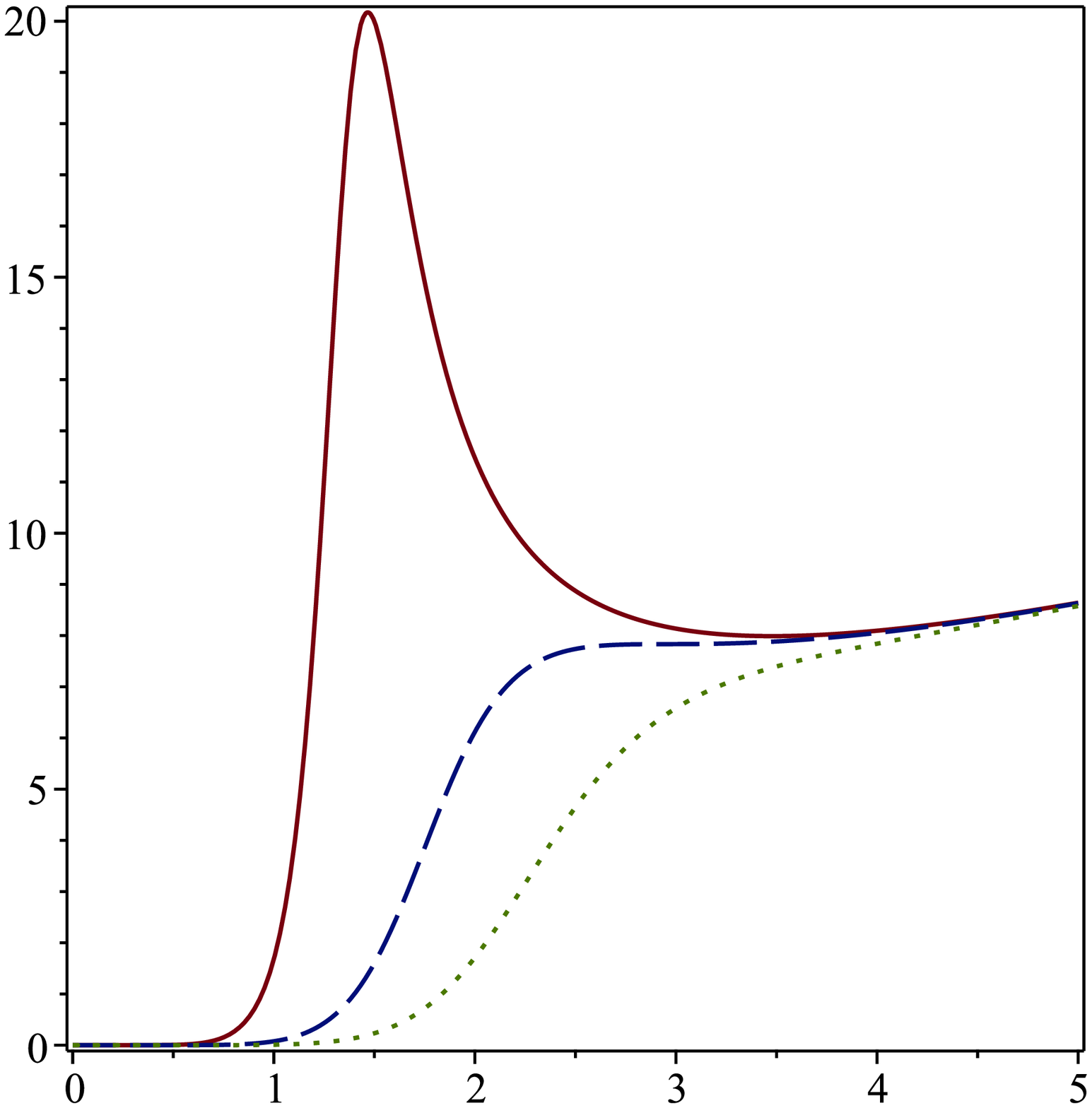} \\
 \tilde r & \tilde r  \\
 &  \\
(c)    &  (d) 
\end{array}
$$	
\caption{  The  specific angular momentum  $\tilde h^2$, as  function of $\tilde r$, 
for  $(a)$ the Hernquist-like model $n=1$, $(b)$ the Plummer-like model $n=2$, 
and the shells $(c)$ $n=3$ and $(d)$ $n=6$, with parameter $\tilde a=1$ (solid curves), $1.5$ 
 (dashed curves)  and $2$ (dotted curves).  }
\label{fig:fig4}
\end{figure}

In Figures  \ref{fig:fig3} -  \ref{fig:fig4}  we show   the curves of tangential speed (rotation curves)  $v_c$ and the specific
angular momentum $\tilde h = h /(GM)$ for  the same value of the   parameters, 
as functions of $\tilde r$.  We observer  that the rotation curves  are not  flat  which means that these configurations of matter     can only  model, as  in the case $n=1$ and $n=2$,   the   visible matter presents in the central regions of the  galaxies such as the  galactic bulge  and the inner core where the relativistic effect can be important and the observed rotation curves have a behavior similar to those shown in figures \cite{Flynn}.
We also see that the speed of the  particles  always is less than  light speed in agreement the dominant energy condition. 
When the parameter $a$ is increased  the motion of the particles 
becomes less relativistic   and  can stabilize  the orbits against radial perturbation.
Thus, except the shell $n=6$  and parameter  $a=1$, these structures  have a physically reasonable behavior.

\section{ Dust disks surrounded by haloes }
Exact solutions of the Einstein field equations  which represent the field of a  disk 
can be obtained by writing  the metric (\ref{eq:met}) in cylindrical coordinates 
($t$, $R$, $z$, $\varphi$)  
\begin{equation}
ds^2 =  -f(R,z)  dt^2 + f(R,z)^{-1} (dR^2 + dz^2  + R^2  d\varphi ^2),  \label{eq:met-cyl}
\end{equation}
and  using  the well known ``displace,
cut and reflect'' method  that was first 
used by Kuzmin \cite{KUZMIN} and Toomre \cite{TOOMRE} to constructed Newtonian
models of disks, and later extended to general relativity 
\cite{BLK,BLP,BL,GL2}. Given a solution of the
Einstein's equations, this procedure is mathematically equivalent to apply
the transformation $z \rightarrow |z| + z_0$,
with $z_0$ constant.  The method is the gravitational version  of the method of images of electrostatic.

The content of matter on the disks can be analyzed using  the 
formalism of distributions in
curved spacetimes  \cite{PH,LICH,TAUB,IS1,IS2}. 
For the metric (\ref{eq:met-cyl}) obtains  a  dust 
(zero pressure) disk with 
 surface energy density \cite{OG}
\begin{equation}
\epsilon = \ \frac{1}{4 \pi G}  f^{-1/2}  f_{,z}   .
\end{equation}

In addition, the tangential  speed $v_c$    and the specific angular momentum $h$
for geodesic motion of test particles on the disks plane are given by 
\begin{subequations}\begin{eqnarray}
v_c^2 &=&  \frac {R f_{,R}} {2f - R f_{,R} }, \\
&  & \nonumber   \\
h^2 &=& \frac{R^2 f^{-1} v_c ^2}{1- v_c ^2}.
\end{eqnarray} \end{subequations} 
 All the quantities are evaluated at $z = 0^+$. 

\begin{figure}
$$
\begin{array}{c}
\includegraphics[width=0.7\textwidth]{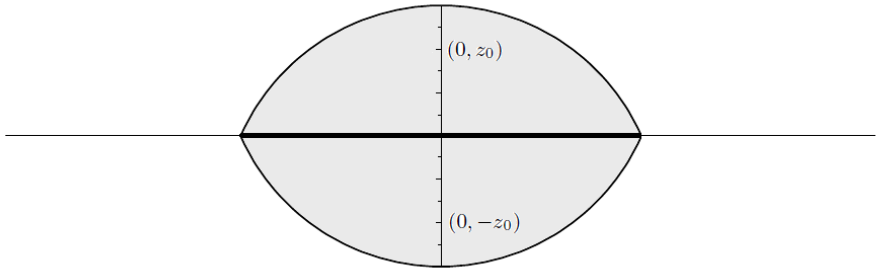}
\end{array}
$$	
\caption{
Schematic representation of the system disk plus halo using  the  ``displace,
cut and reflect'' method. The points  $(0,-z_0)$  and   $(0,z_0)$ 
are the centers of spheres corresponding to the spherical caps located  above and below of the disk, respectively.}
\label{fig:fig5}
\end{figure}

Now when this procedure is applied to the above  structures with   $0 \leq z_0 \leq r_c$,
one obtains  at points with  $z> 0$  a matter spherical  cap  corresponding to  a sphere   with center located at point  $(R,z) =(0,-z_0)$,  and when $z<0$ a matter  cap of a sphere with    center at $(R,z) =(0,z_0)$.  The discontinuity in  the normal derivative of the metric tensor through the plane  $z=0$ introduces a planar matter distribution   which  is interpreted as the field of a thin disk.  The result in this case  is  a system composite by dust disks surrounded by a halo of matter  (figure \ref{fig:fig5}).

For the potential-density pair   (\ref{eq:plu1}) and (\ref{eq:plu2}) we have 
\begin{equation}
f =  \left [   1 +  \left (  [ \tilde R^2 + (|\tilde z| + \tilde z_0 )^2   ]^{n/2}  
+ \tilde a^n  \right )  ^{-1/n} \right ] ^{-4}    \label{eq:potential}  ,
\end{equation}
where $\tilde R = R/(GM)$,  $\tilde z= z/(GM)$,  $\tilde z_0= z_0/(GM)$,  $\tilde a = a/(GM)$,  and  the surface density energy is
\begin{equation}
\epsilon =  \frac {4  \tilde z_0  \bar r  ^{n-2} (\bar r ^n + \tilde a^n)^{2/n - 1}  }{ \pi G     [2  (\bar r ^n + \tilde a^n) ^{1/n} +1 ] ^3   },
\end{equation}
where $\bar r = \sqrt {\tilde R ^ 2 + \tilde a ^2}$.

\begin{figure}
$$
\begin{array}{ccc}
\tilde \epsilon &  v_c ^2 &\tilde h^2   \\
\includegraphics[width=0.3\textwidth]{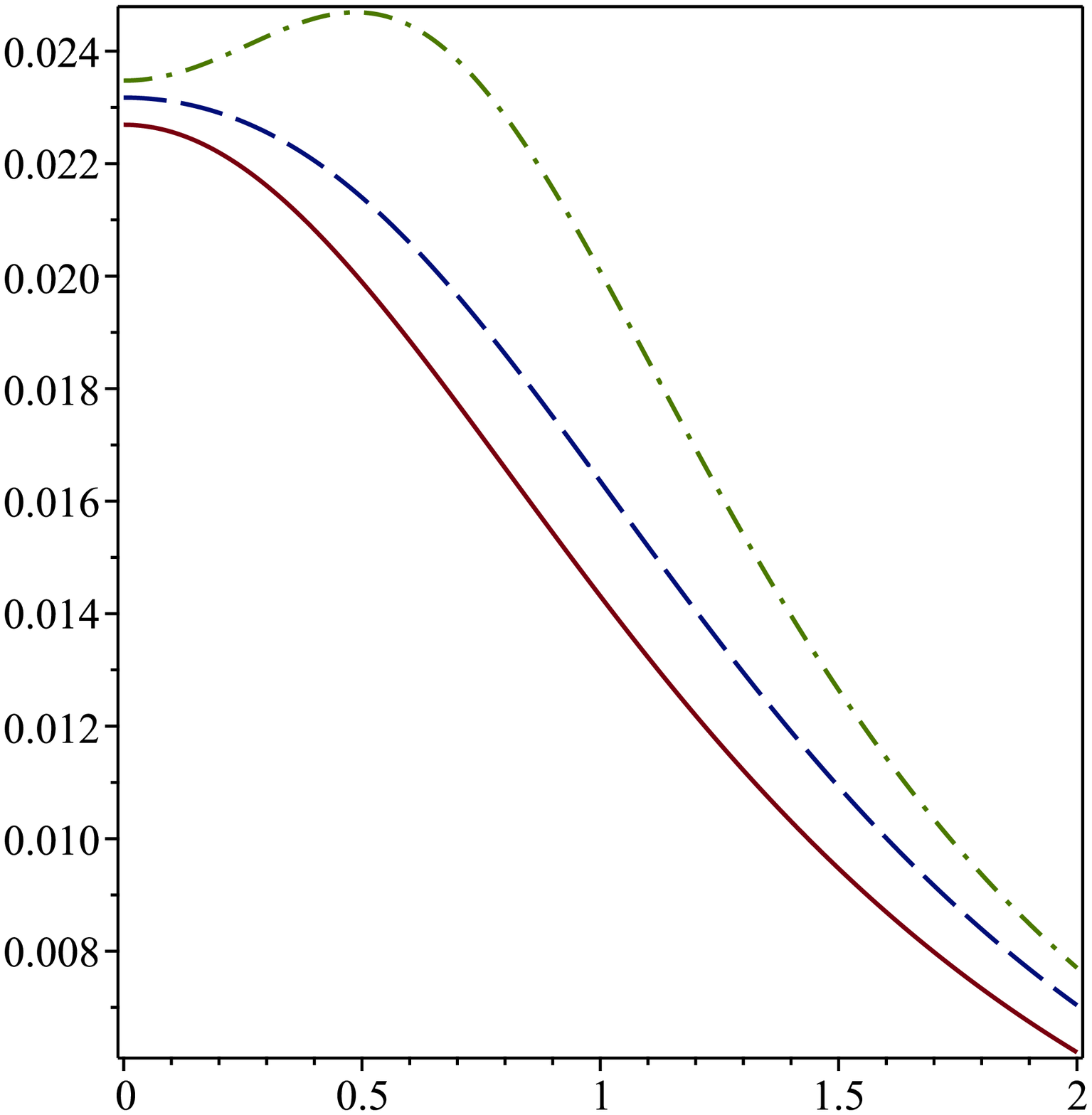} &      
\includegraphics[width=0.3\textwidth]{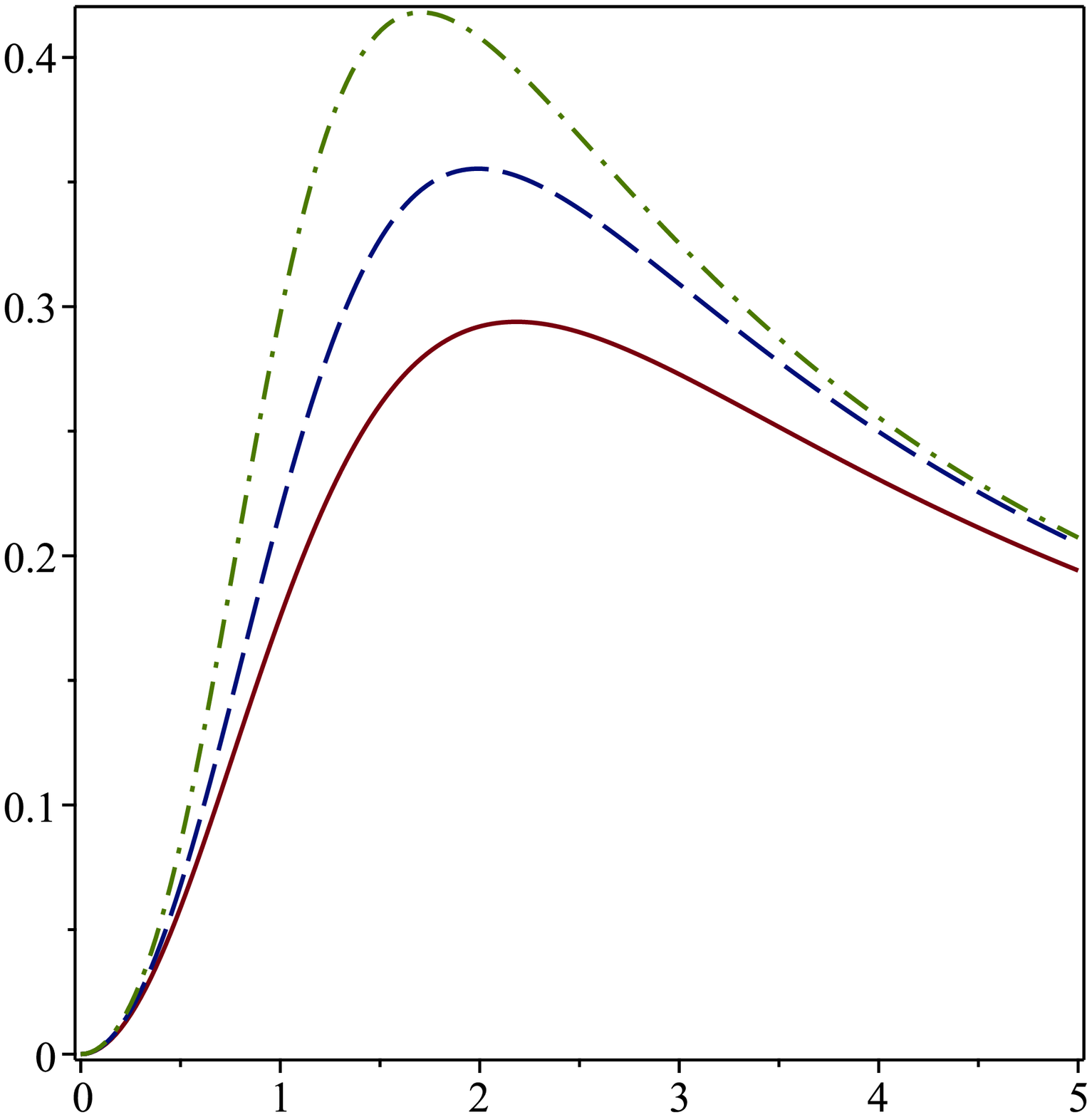} &
  \includegraphics[width=0.3\textwidth]{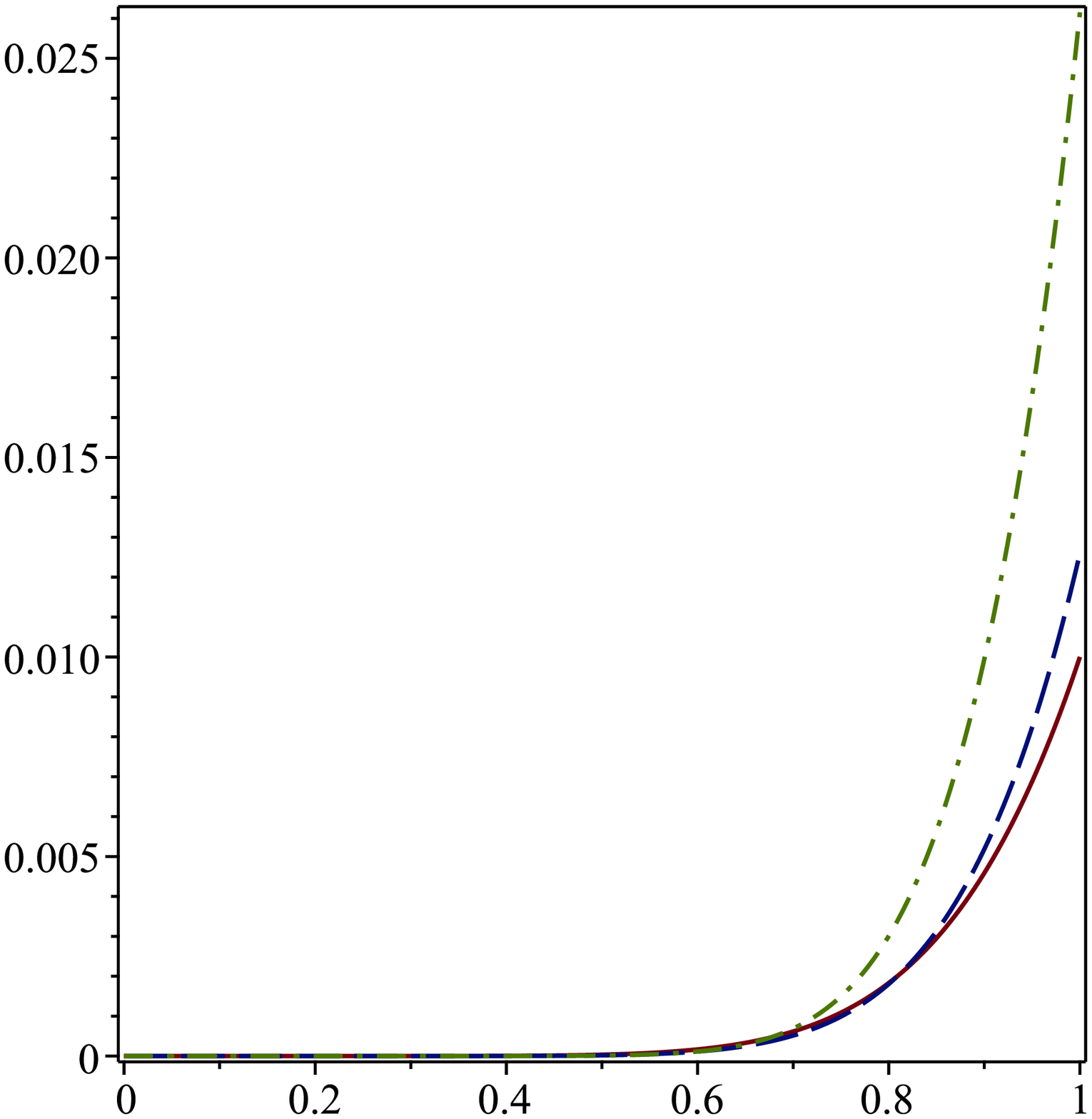}   \\  
 \tilde R &  \tilde R  &\tilde R  \\
&& \\
(a)    &  (b) & (c)  
\end{array}
$$	
\caption{ $(a)$ The  surface energy density $\tilde \epsilon$, $(b)$ the tangential speed (rotation curves)  $v_c$ and   $(c)$ the specific
angular momentum $\tilde h$  for the relativistic disk with 
$n=2$ (solid curves),  $3$ (dashed curves), $6$ (dash-dotted  curves)  and parameters  $\tilde a =\tilde z_0=1$,  as  functions of $\tilde R$ }
\label{fig:fig6}
\end{figure}

In Fig. \ref{fig:fig6}   we show    the  surface energy density 
$\tilde \epsilon = G \epsilon$,
the tangential speed   $v_c$ and the specific
angular momentum $\tilde h = h /(GM)$ for disks  with  $n=2$, $3$,  $6$  and parameters $\tilde a =\tilde z_0=1$,  as  functions of $\tilde R$.  
We  see that the energy is everywhere positive, its maximum occurs in
the center of the disk, and it vanishes sufficiently fast as $\tilde R$ increases. When the parameter 
$n$ is increased, the energy density increases.
We observer that
circular speed of particles is always a quantity less than the speed of light and that 
the increasing of $n$   makes  more  relativistic the  orbits. 
We also  see  that the rotation curves  are not  flat  which means that these  structures     can only  model the   luminous matter of  galaxies.
In addition, in all cases   the motion of particles is stable  against radial
perturbation. The same behavior  is observed for other 
 values of parameter. 

\section{Discussion}
 Spherical shell models for a Majumdar-Papapetrou type  conformastatic  spacetime was construct,  using as  seed  a Newtonian potential-density pair.
The  formalism  presented allows to construct   different  relativistic spherical configurations  of matter  such as spheres and   shell-like configurations. 
The method  was  illustrated 
in the case of   a  family  of Plummer-Hernquist type relativistic  thick spherical shells.  
Besides shells, this potential-density pair
 also  permits to construct   configurations of matter
with a central cusp  and with a finite central density. 
These structures  were  then used to study  models of 
dust disks  surrounded by a halo  of matter.
The models  considered  satisfied  all the energy conditions.


\newpage



\end{document}